\documentclass[prl,showpacs,amsmath,twocolumn,floatfix]{revtex4}

\usepackage{color}

\usepackage{graphicx}
\usepackage{amsmath}
\usepackage{bm}

\def\beqn{\begin{eqnarray}}
\def\eeqn{\end{eqnarray}}
\def\barr{\begin{array}}
\def\earr{\end{array}}
\def\btab{\begin{tabular}}
\def\etab{\end{tabular}}
\def\bite{\begin{itemize}}
\def\eite{\end{itemize}}
\def\bcen{\begin{center}}
\def\ecen{\end{center}}

\def\eq{\begin{equation}}
\def\ee{\end{equation}}
\def\nn{\nonumber}

\def\q2dagger{q_2\hspace{-0.35cm}/\;}

\begin{document}
\title{Low energy theorem for virtual Compton scattering and generalized sum 
rules of the nucleon}
\author{Mikhail Gorchtein and Adam P. Szczepaniak}
\affiliation{Physics Department and Nuclear Theory Center, \\
Indiana University, Bloomington, IN 47403}
\begin{abstract}
We formulate the low energy theorem for virtual Compton scattering off a 
nucleon and examine its consequences for generalized nucleon polarizabilites. 
As a result of a new, model independent definition of the low energy limit for 
VVCS reaction, all generalized sum rules of the nucleon have continuous limit 
for real photons and obtain 
contributions from the $t$-channel that were not included previously. 
\end{abstract}
\maketitle
Understanding proton  structure remains to be one of the top priorities in 
physics of strong interactions. The quest  started in the 30's with the first 
measurement of magnetic moment of the nucleon ~\cite{stern} that  was found 
to be anomalously large indicating that proton is not a point-like Dirac 
particle. In the late 50's, application of dispersion relations to real 
Compton scattering together with  a low energy theorem (LET) gave rise to 
nucleon sum rules. The sum rules express a model-independent correspondence 
between static properties of the nucleon and photo-absorption spectrum
~\cite{low}. In particular, the Gerasimov-Drell-Hearn (GDH) sum rule relates 
unambiguously the 
value of the anomalous magnetic moment (a.m.m.) of the nucleon $\kappa$ to the 
spin-dependent cross-section ~\cite{gdh}, 
$-e^2\kappa^2/2M^2= (1/\pi) \int d\nu (\sigma_{1/2}-\sigma_{3/2})/\nu$,
with $e$ the electric charge, $M$ the nucleon mass,  
and the helicity cross sections $\sigma_{1/2}(\sigma_{3/2})$ 
refer to total photo-absorption $\gamma+N\to X$ of circularly polarized photons 
by polarized nucleon with the helicities parallel (antiparallel) to each other, 
respectively. 
The validity of the GDH sum rule was proven analytically in QED for the 
electron that receives  a non-zero a.m.m.,  $\alpha_{em}/2\pi$ from  
radiative corrections ~\cite{gdh_qed}. Unfortunately, such a consistency check 
within QCD is not possible, and it is important to have a direct comparison of 
photo-absobtion nucleon sum rule  with the experimental data. Such data were 
obtained  by the GDH collaboration at MAMI ~\cite{gdh_mainz} that 
confirmed its validity within experimental precision. \\
\indent
A more detailed study of the internal structure of the nucleon has become 
possible with electron scattering experiments 
where the  electromagnetic  interaction is mediated by a virtual photon. 
Elastic electron scattering probes spatial  distribution of charge and 
magnetization inside the nucleon~\cite{elastff}. 
Further insight into nucleon structure came from deep inelastic 
electron scattering (DIS) experiments at SLAC ~\cite{SLAC} that 
confirmed that quarks and gluons are the building blocks of the nucleon. 
Having observed the intimate connection between static properties of the 
nucleon and the photo-absorption cross section, it is logical to expect a 
similar relation between form factors and absorption cross section for 
virtual photons. \\
\indent
In ~\cite{marcbarbaradrechsel}, an attempt to obtain such generalized 
sum rules was made by considering the forward limit of symmetric virtual 
Compton scattering (VVCS) 
with equal virtualities of the two (spacelike) photons $Q^2 = -q^2 = -q'^2$. 
Applying analyticity and unitarity 
to the four independent amplitudes describing this process, their imaginary 
parts were related to the DIS structure functions. 
Their real parts are obtained from dispersion relations at fixed $Q^2$ and 
fixed $t=0$. According to the idea of LET, these real 
parts are expanded into Taylor series in variable $\nu$, thus reducing the 
problem to a dispersion representation of the low energy coefficients of the 
expansion 
(constants in $\nu$ and functions of $Q^2$). The lowest order coefficients, 
on the other hand, can be calculated directly from the nucleon-pole graphs and 
thus related to the elastic form factors. However, in
~\cite{marcbarbaradrechsel}, 
these constants were obtained at such a kinematical point where the virtual 
photon brings the nucleon off its mass shell, and where no information on the 
form factors is available. As a result, the low energy limit of the 
spin-independent VVCS amplitude does not match with that for real photons, 
the well-known Thomson term. The situation is even worse for the spin-dependent 
part: if taken at finite $Q^2$, its low energy limit according 
to~\cite{marcbarbaradrechsel} has a singularity at the real photon point. 
Correspondingly, the GDH sum rule as function of $Q^2$ is in general not 
defined and only exists for real photons, and the only generalization that is 
achieved concerns the GDH integral.

The main goal of this letter is to show how these problems are circumvented 
when low-energy  
expansion is set up around the correct point in the Mandelstam plane of 
kinematical variables  and derive a new set of sum rules for the low energy 
parameters. 

\begin{figure}[h]
{\includegraphics[height=3cm]{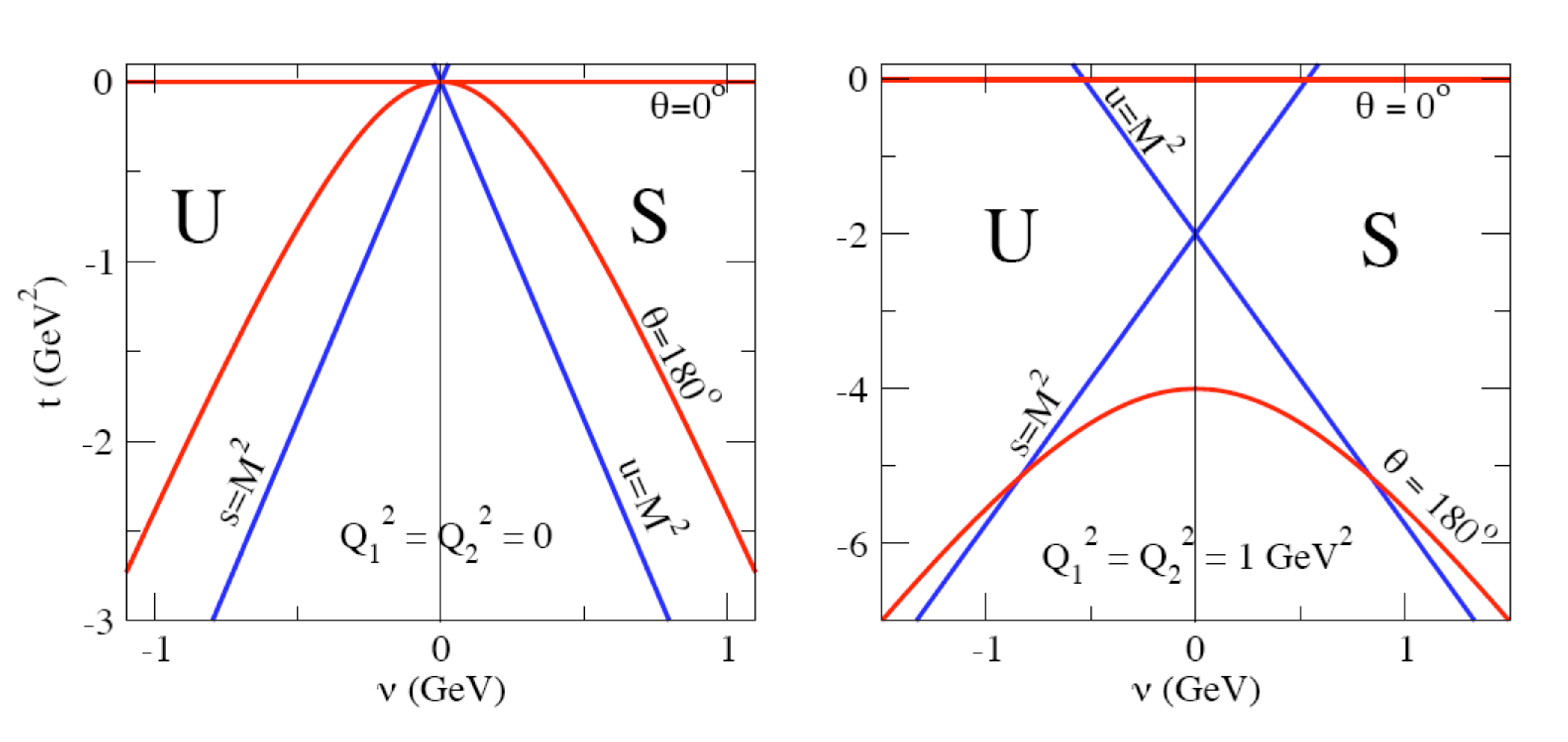}}
\caption{ The Mandelstam $\nu,t$ plane of real Compton scattering (RCS) 
(left panel) and  symmetric VVCS (right panel). 
The physical region is bound by the $\theta=0^0,180^0$ lines. The nucleon 
pole in the Born $s$ ($u)$ channel amplitudes is located along the 
$s(u) = M^2$ lines. The $t=0$ and $\theta=0^0$ lines coincide. }  
 \label{fig:mandelstam}
\end{figure}

\begin{figure}[h]
{\includegraphics[height=3cm]{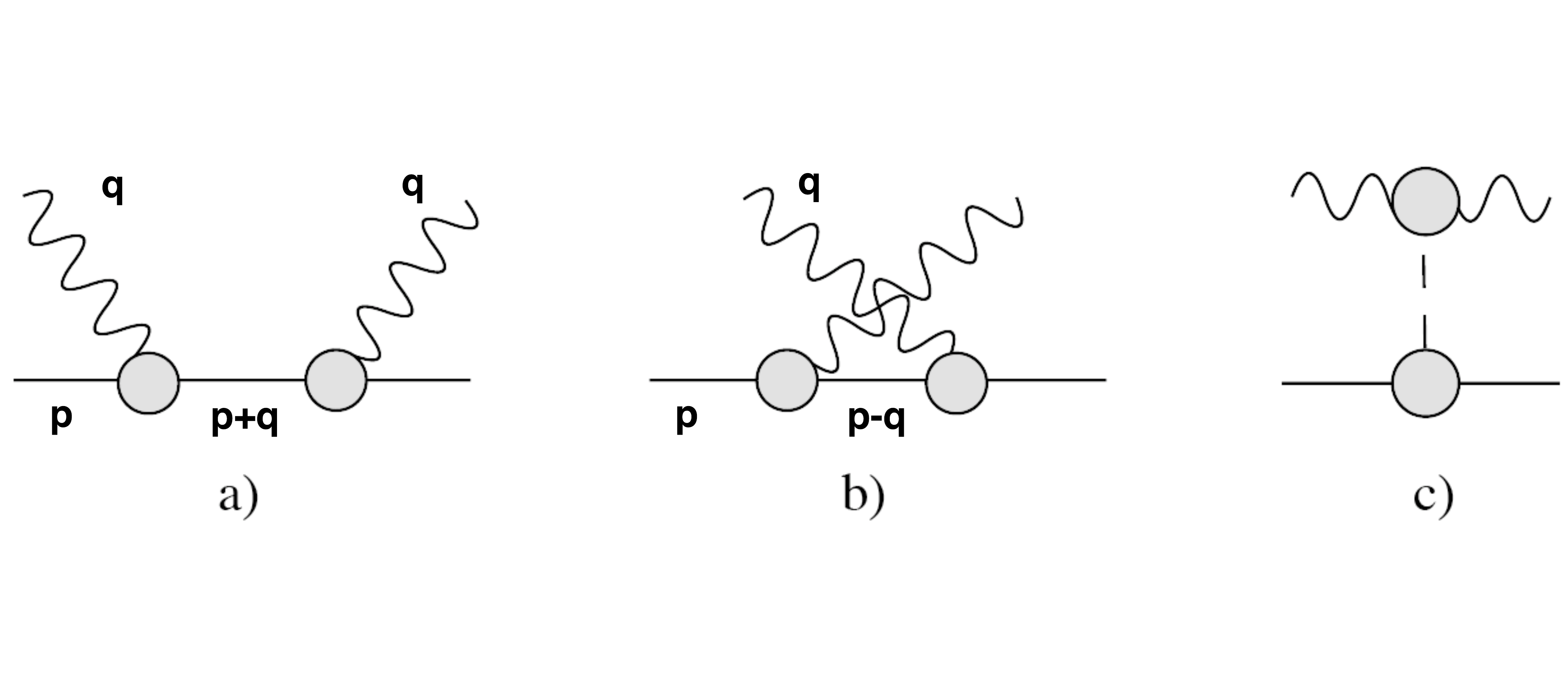}}
\caption{The generalized Born contributions to Compton scattering: 
nucleon exchange diagrams  (a,b) and the $t$-channel $\pi^0$-exchange 
(c). The blobs denote form factors.}
\label{fig:born}
\end{figure}

The twelve invariant amplitudes describing symmetric VVCS 
are functions of three independent  Mandelstam  variables, which are 
conveniently chosen as $\nu = (s-u)/4M$, $t$, and $Q^2$ where, 
$s=(p+q)^2$, $u=(p-q')^2$, $t=(p'-p)^2$ satisfy 
$s + t + u + 2Q^2= 2M^2$   with $M$ being the nucleon mass. 
Since the $s$ channel,  $\gamma^*(q)+N(p)\to\gamma^*(q')+N(p')$, and 
$u$ channel, $(q \leftrightarrow -q'$) represent  the same physical process, the
individual amplitudes have well defined parity under $\nu \to -\nu$ exchange, 
which  makes  the variable $\nu$ particularly useful. 
For fixed $Q^2$, the physical region of the $s$-channel reaction is  
fixed by the following conditions: {\it i)} the total energy is above 
the threshold for the reaction in the $s$-channel,  
while {\it ii)} it is below physical thresholds in $u$- and $t$-channels,
and {\it iii) }  the cosine of the scattering angle takes physical values, and 
similarly for the $u$-channel. 
The Mandelstam $\nu-t$ plane is shown in Fig.~\ref{fig:mandelstam}. 
The $t$-channel physical region lies at 
$t\geq4M^2$ and is not shown here. \\
\indent
We next consider the analytical structure of a   VVCS amplitude. It has 
two distinct classes of contributions: the (generalized) Born contributions 
that correspond to a single-particle exchange in the $s,u$ or 
$t$-channel,  shown in Fig.\ref{fig:born}, 
and the remaining, non-Born contributions  that are associated with  two or 
more particle intermediate states in either channel.
These two contributions have different analytical structure, so they should 
be treated separatly. Furthermore, away from the  physical nucleon point, 
$s=u=M^2$, the Born amplitude, however, contains off-mass shell particles and  
have to be treated  with care to avoid model dependence. 
For the general case  $Q_1^2 \ne Q_2^2$  we can summarize these observations 
in the following  statement: 
{\it Low Energy Theorem for VVCS: 
The low energy  expansion for the general VVCS 
reaction is defined  in the Mandelstam $\nu,t$ plane around the  
$\nu =  0, \;q\cdot q' = -(t + Q_1^2 + Q_2^2)/2  = 0$ point which automatically 
corresponds to $s=u=M^2$. At this point, 
the contribution from the nucleon-exchange, Born amplitude is unambiguous given in terms of 
physical observables and it can be isolated. 
The generalized polarizabilities parametrize the leading terms of the 
low energy expansion of the rest amplitude in powers of $\nu$.
} \\
Note that in Breit frame defined by $\vec{p}+\vec{p'}=0$ the low energy limit 
defined above corresponds to vanishing of both photon energies since 
$\omega=\omega'=M\nu/\sqrt{M^2-t/4}$, independently on the 
virtualities.
In  contrast,  in  \cite{marcbarbaradrechsel}, 
the low-energy expansion was formulated 
as expansion in $\nu$ around the point $\nu=0,t=0$ instead, and this lead to 
the controversial results discussed earlier. 
In the following, we will use the $\nu=q\cdot q' = 0$ point to 
define the low energy expansion and rederive sum rules for nucleon 
polarizabilities.
We will demonstrate that once the correct point for low energy expansion is 
chosen, the generalized Thomson term is obtained correctly {\it i.e.} the 
VVCS amplitude reduces to that for RCS in the limit $Q^2 \to 0$. 
To use the power of  dispersion relations in VVCS it is necessary to  introduce 
a set of scalar amplitudes, $F_i(\nu,t,Q^2)$  defined as coefficients in  
expansion of the hardonic tensor $T^{\mu\nu}$ in a basis of twelve 
independent tensors $T^{\mu\nu}_i$ that are gauge-invariant and free from 
kinematical singularities and constraints, \cite{tarrach,metz},
\beqn
\bar{u}(p')T^{\mu\nu}u(p)\,=\,\sum_i\bar{u}(p')T_i^{\mu\nu}u(p) F_i(\nu,t,Q^2)
\label{basis_old}
\eeqn
In the above sum the label $i$ takes on  twelve values 
$i=1$,$2$,$4$,$6$,$7$,$8$,$10$,$12$,$17$,$18$,$19$,$21$ - using the original 
labels from \cite{tarrach,metz} where the general $(Q_1^2 \ne Q_2^2)$ VVCS 
amplitude was considered.   In the forward limit,  $t=0$, only four out of the 
twelve tensors   are independent   and can be chosen to coincide with 
$T^{\mu\nu}_i$ for  $i=1$,$2$,$7$,$17$.  From  the remaining eight tensors 
$j= 4$,$6$,$8$,$10$,$12$,$18$,$19$,$21 \ne i$  we extract the components  
$\tilde{T}_j^{\mu\nu}$ that vanish in the forward direction,    
$\tilde{T}_j^{\mu\nu}=T_j^{\mu\nu}-\sum_{i=1,2,7,17\ne j}c_{ij}T_i^{\mu\nu}$
and write the  VVCS amplitude as
$T^{\mu\nu}=\sum_{i}f_i T_i^{\mu\nu} +\sum_{j}F_j \tilde{T}_j^{\mu\nu}$. 
The first term, with  $f_i=F_i+\sum_jc_{ij}F_j$  contains  the four tensors 
that  are independent in the forward limit and the second term contains the 
remaining  eight tensors that complete the basis  for $t \ne 0$ and  
vanish in the forward limit. 
Our four forward VVCS amplitudes $f_i=f_i(\nu,t,Q^2)$ are related to the
amplitudes $T_{1,2},S_{1,2}$  of  \cite{marcbarbaradrechsel} by
\beqn
\hat f_1 & \equiv &  f_1=-\frac{1}{2M}\frac{1}{Q^2}\left[T_1-\frac{1}{2x}T_2\right]\nn\\
\hat f_2  & \equiv  & \nu^2 f_1 =\frac{1}{2M}\frac{1}{4M^2}\frac{1}{2x}T_2\nn\\
\hat f_7 & \equiv  & \nu f_7 =\frac{1}{8M^3}S_1\nn\\
\hat f_{17} & \equiv &  \nu f_{17} =-\frac{\nu}{4M^3}S_2 \label{low}
\eeqn
As discussed earlier, $s$ and $u$ channel nucleon exchange is unambiguous at 
the low energy, $\nu=0,t=-2Q^2$ point and given by  
$\hat f_i^B(Q^2) = \lim_{\nu \to 0}  \hat f_i^B(\nu,t=-2Q^2,Q^2)$,
\beqn
\hat f^B_1(Q^2) &=&-\frac{e^2}{4M^3}F_P^2\label{eq:born}\\
\hat f^B_2(Q^2) &=&-\frac{e^2}{4M^3}F_D^2\nn\\
\hat f^B_7(Q^2) &=&-\frac{1}{8M^3}F_P^2\nn\\
\hat f^B_{17}(Q^2) &=&\frac{1}{4M^2}\left[(F_D+F_P)^2+\frac{Q^2}{8M^2}F_P^2\right]\nn
\eeqn
where $F_{D,P}(Q^2)$ stand for the Dirac and Pauli elastic form factors of the 
nucleon ( instead of the usual notation $F_{1,2}$ to avoid confusion with the 
DIS structure functions).  Note that residues of the poles in the $s$ and $u$ 
channel vanish at $t=-2Q^2$ so $\hat f^B_i$ are finite. The $\nu^{-1}$, 
$\nu^{-2}$ singularities in the Born amplitudes in Eq.(\ref{eq:born})
$f^B_2,f^B_7,f^B_{17}$ are  cancelled in the full amplitude $T^{\mu\nu}$ by 
corresponding zeros in $T^{\mu\nu}_i$, for $i=2,7,17$. Finally, we notice for 
completeness that the pion-pole 
of Fig.~\ref{fig:born}, (c) does not contribute to any of the four amplitudes. 
It is instructive  to compare these results with the well-known results for 
real Compton scattering. The spin-independent real Compton amplitude is known 
to be finite at zero 
energy and given by the classical Thomson term  
$\bar u(p) T^{\mu\nu}(\nu=0,t=0,Q^2=0) u(p) =2e^2g^{\mu\nu}$~\cite{thompson}.  
Using the tensors $T^{\mu\nu}_i$ from \cite{tarrach,metz} and picking the 
term $\sim g^{\mu\nu}$, we obtain the correct answer when taking $Q^2=0$ limit. 
 
The amplitudes 
$\hat f_i$ defined above are free from kinematical singularities and can 
be used in  dispersion relation  calculation. 
All four amplitudes $\hat f_i$ are even functions of $\nu$ and as a result 
obey a dispersion relation at fixed-$t$ and $Q^2$  in the form
\beqn
{\rm Re}\hat f_i(\nu,t,Q^2)\,=\frac{1}{\pi}\int_{\nu^2_0}^\infty
\frac{d\nu'^2}{\nu'^2-\nu^2}{\rm Im}\hat f_i(\nu',t,Q^2),
\label{eq:dr}
\eeqn
 unless a subtraction is needed to ensure the convergence of the integral.
In the forward direction, $t=0$, the optical theorem relates imaginary 
parts to the DIS structure functions,
\beqn
{\rm Im}\hat f_1(\nu,0,Q^2)&=&-\frac{\pi e^2}{MQ^2}
\left[F_1(x,Q^2)-\frac{1}{2x}F_2(x,Q^2)\right]\nn\\
{\rm Im}\hat f_2(\nu,0,Q^2)&=&\frac{\pi e^2}{4M^3}\frac{1}{2x}F_2(x,Q^2)\nn\\
{\rm Im}\hat f_7(\nu,0,Q^2)&=&\frac{\pi e^2}{4M^2\nu}g_1(x,Q^2)\nn\\
{\rm Im}\hat f_{17}(\nu,0,Q^2)&=&-\frac{\pi e^2}{2M\nu}g_2(x,Q^2),
\eeqn
where the usual Bjorken variable $x=Q^2/2M\nu$ was introduced.
To obtain the generalized sum rules, the above forward dispersion relations 
have to be connected to the 
$t=-2Q^2$ point where the separation of the VVCS amplitude into Born and 
non-Born parts is well defined. This can be done by invoking analyticity 
in the $t$-channel and it  leads to 
subtracted dispersion relations in $t$ for each of the four amplitudes,
\beqn
&&{\rm Re}\hat f_i(\nu,-2Q^2,Q^2)-{\rm Re} \hat f_i(\nu,0,Q^2)\\
&=&
-\frac{2Q^2}{\pi}\int \frac{dt'}{t'(t'+2Q^2)}
{\rm Im}\hat f_i(\nu,t',Q^2)\nn \label{t} 
\eeqn
This $t$-channel contribution does not affect the real photon point since for 
real photons the low energy expansion is defined at $t=0$ but it is 
non-zero at finite $Q^2$. 
Finally, we expand the unknown non-Born part of the amplitudes around the 
point $\nu=0$ at $t=-2Q^2$. 
To accomplish this, we rewrite the VVCS tensors 
in terms of electric and  magnetic fields of the initial and final photon. 
We next define the field strength  
tensors , $F^{\mu\nu}=-ie(q^\mu\epsilon^\nu-q^\nu\epsilon^\mu)$ 
and $F'^{\mu\nu}=-ie(q'^\mu\epsilon'^\nu-q'^\nu\epsilon'^\mu)$. 
As discussed earlier, it is the Breit frame, where energies of the initial 
and final photon vanish simultaneously at the nucleon point.
Then, the spin-independent tensors can be rewritten through the electromagnetic 
fields in Breit frame as
\beqn
e^2\varepsilon_\mu\varepsilon'^*_\nu 
T_1^{\mu\nu}&=&-\frac{1}{2}F^{\mu\nu}F'^*_{\mu\nu}
\,=\,\vec{E}\cdot\vec{E}'^* - \vec{B}\cdot\vec{B}'^* \\
e^2\varepsilon_\mu\varepsilon'^*_\nu 
T_2^{\mu\nu}&=&-4(P_\mu F^{\mu\alpha})(P^\nu F'^*_{\nu\alpha})\,=\,
(4M^2-t) \vec{E}\cdot\vec{E}'^* \nonumber
\eeqn
\indent
The dipole polarizabilities quantify the electric (magnetic) dipole induced in 
low-energy external field $\vec{E}(\vec{B})$ in the direction of the field, 
$\vec{d} =4\pi\alpha\vec{E}$, $\vec{m} =4\pi\beta\vec{B}$. 
These dipoles then interact with the outgoing photon electric and magnetic 
fields, respectively.  
Approximating the response of the internal structure of the nucleon by 
the linear form, we arrive to the low energy expansion of the 
spin-independent amplitudes, $\hat f_{1,2}(\nu,-2Q^2,Q^2)$,  
\beqn
\hat f_1 &=&-\frac{e^2}{4M^3}F_P^2-4\pi\beta(Q^2)+O(\nu^2)
\label{eq:lex_nospin}\\
\hat f_2 &=&-\frac{e^2}{4M^3 }F_D^2 + \frac{4\pi \nu^2}{4 M^2 + 2Q^2} \left[ \alpha(Q^2)+\beta(Q^2)\right] 
\nonumber \\ & + & O(\nu^4)\nn 
\eeqn
We notice that for low energy virtual photons in Breit frame, only the 
transverse components are small, $\vec{E}_T=\omega\vec{\varepsilon}_T =O(\nu)$, 
while 
the longitudinal electric and (transverse) magnetic components are finite, 
$E_L,B\sim|\vec{q}|\sim\sqrt{Q^2}$. 
Because of the choice of the kinematical point $t=-2Q^2$, however, 
the longitudinal electric fields in the spin-independent part contribute
at order $O(\nu^2)$: 
$(\vec{E}_L\cdot\vec{E}'^*_L)\sim(\vec{q}\cdot\vec{q'})=\omega^2 = O(\nu^2)$. 
This explains, why contribution from  polarizabilities to $\hat f_2$ is of 
relative 
order $\nu^2$ with respect to the Born contribution, while the contribution 
of the magnetic polarizability to $\hat f_1$ appears at the same order in energy
as the Born contribution. The low energy expansion of Eq.(\ref{eq:lex_nospin}) 
is a new result. 
We contrast this with the result of \cite{marcbarbaradrechsel} where 
the low  energy expansion of the amplitudes $f_{1,2}$ in the forward direction 
is in terms of the polarizabilities $\alpha_L$ and  $(\alpha+\beta)$ instead of 
$\beta$ and $(\alpha+\beta)$, respectively with both expansions starting at relative order 
$\nu^2$ obtained by a mere analogy with real Compton scattering. Our results 
demonstrate that this analogy may be misleading. 

Finally we write down the generalized sum rules for the low energy 
constants appearing the expansion of all four amplitudes $\hat f_{1,2,7,17}$. 
Using Eqs.~\ref{low},\ref{eq:dr},~\ref{t} for the first three amplitude we 
find,
\begin{widetext} 
\beqn
\frac{e^2}{4M^3}F_P^2 
+  4\pi\beta(Q^2) &= & \frac{2e^2}{MQ^2}
\int_0^{x_\pi}\frac{dx}{x}\left[F_1(x,Q^2)-\frac{1}{2x}F_2(x,Q^2)\right] 
+\frac{2Q^2}{\pi}\int  \frac{dt'}{t'(t'+2Q^2)}
{\rm Im}\hat f_1(0,t',Q^2)\nn\\
4\pi\left[\alpha(Q^2)+\beta(Q^2) \right]  &= & \frac{4e^2M}{Q^4} \left[ 1 + \frac{Q^2}{2M^2} \right] \int_0^{x_\pi}dx
F_2(x,Q^2) -\frac{2Q^2}{\pi}\int  dt'
\frac{1}{t'(t'+2Q^2)}  \frac{d}{d\nu^2} {\rm Im}\hat f_1(0,t',Q^2) 
  \nonumber \\
-\frac{e^2}{8M^3}F_P^2(Q^2) &= & \frac{e^2}{MQ^2}\int_0^{x_\pi}dxg_1(x,Q^2)-\frac{2Q^2}{\pi}\int \frac{dt'}{t'(t'+2Q^2)} {\rm Im}\hat f_7(0,t',Q^2)
\eeqn
\end{widetext}
where the upper limit of integration $x_\pi$ corresponds to the threshold of 
pion production, $x_\pi=Q^2/(Q^2+m_\pi^2+2Mm_\pi) <1$.  
The integrals over $t$ receive  contributions from unitarity in the 
$t$-channel starting with the reaction $\gamma\gamma\to\pi\pi\to N\bar{N}$ 
at $t\geq(2m_\pi)^2$, and from the $su$ spectral region at negative $t$.
The sum rule for the amplitude $\hat f_{17}$ gives rise to the 
Burkhard-Cotingham sum rule, $\int_0^1g_2(x,Q^2)dx=0$. This sum rule is not 
related to the low energy expansion of the corresponding  amplitude, and that 
is why we dropped it here. \\
\indent
The first sum rule that give a  dispersion  relation representation for  the magnetic 
polarizability $\beta$ is new. Due to the structure of the tensor $T_1$, 
$T_1=-(qq')g^{\mu\nu}+q^\nu q'^\mu$ at $(qq')=0$, the result is purely 
magnetic, whereas at $t=0$ it has a longitudinal component that led the authors 
of \cite{marcbarbaradrechsel} to the sum rule for $\alpha_L$. The convergence 
of this sum rule is however unclear. 
The second sum rule is the generalization of Baldin sum rule to which it 
reduces at $Q^2=0$. As compared to the corresponding generalization from 
\cite{marcbarbaradrechsel} we notice (besides the $t$-channel piece) the 
relative factor of $1+Q^2/2M^2$  
whose origin is in the usual Breit frame factor $(p^0)^2=M^2-t/4$ entering the 
low energy expansion of $\hat f_2$ instead of $M^2$ in the lab frame.
The third sum rule is the generalization of the GDH sum rule. 
At the real photon point it has a 
negative value, whereas the high $Q^2$ data for the first moment of $g_1$ 
suggest a small positive value. The transition between the two regimes is 
a subject of intensive experimentally studies~\cite{JLab-GDH}.  Our 
results suggest that at finite $Q^2$ the sum rule has the $t$-channel 
contribution that is missing in previous analyses. Due to small pion mass, it is 
expected to lead to a fast variation of this contribution at low $Q^2$. 
The indications of this behaviour were obtained in HBChPT \cite{chpt}
although the range of 
validity of those calculations is constrained to very low values of $Q^2$. 
We leave the computation of the $t$-channel contribution to an upcoming work. 
This work was supported in part by the US Department of Energy grant under 
contract DE-FG0287ER40365, by the US National Science Foundation under grant 
PHY-0555232. 


\begin{thebibliography}{9}
\bibitem{thompson}  The more familiar result is $-\alpha_{em}/M$ and it 
corresponds to the Compton normalization that differs from ours by a factor of 
1/2M coming from the nucleon spinors and 
$1/4\pi$ due to the Gaussian units of 
the electric charge, $e_G^2=\frac{e^2}{4\pi}$ that is conventionally used in 
classical electrodynamics.
\end{thebibliography}
\end{document}